\documentclass[aps,amsmath,twocolumn,amssymb,floatfixng,showpacs, superscriptaddress,footinbib, longbibliography]{revtex4-1}

\usepackage{graphicx}% Include figure files
\usepackage{dcolumn}% Align table columns on decimal point
\usepackage{multirow}
\usepackage{booktabs}
\usepackage{bm,color}
\usepackage{braket}
\usepackage{amsmath,amssymb}
\usepackage[colorlinks,linkcolor=blue,hyperindex,CJKbookmarks]{hyperref}
\usepackage{epstopdf}

\def\x1{{\it x}}
\def\y1{{\it y}}
\def\z1{{\it z}}

\def\i{{\it i}}

\def\be{\begin {equation}}
\def\ee{\end {equation}}
\def\ber{\begin {eqnarray}}
\def\eer{\end {eqnarray}}
\def\bers{\begin {eqnarray*}}
\def\eers{\end {eqnarray*}}

\newcommand{\bs}[1]{\textcolor{black}{#1}}

\begin{document}
\title{\bs {Large bulk photovoltaic effect and Fermi surface mediated its enhancement with chemical potential in ZnGeP$_2$} }
\author{Banasree Sadhukhan}
\affiliation{Department of Physics and Nanotechnology, SRM Institute of Science and Technology, Kattankulathur, 603203, Chennai, Tamil Nadu, India}
\email{banasres@srmist.edu.in}
\affiliation{Tata Institute of Fundamental Research, Hyderabad, Telangana 500046,  India}

\begin{abstract}

Bulk photovoltaic effect is a non-linear response in noncentrosymmetric materials that converts light {\bs{into}} DC current.  In this work, we investigate the optical linear and non-linear responses in a chalcopyrite semiconductor ZnGeP$_2$.  {\bs {We report large bulk photovoltaics namely shift and circular photogalvanic current conductivities which are 4.46 $\mu$A/V$^2$ and -5.49 $\mu$A/V$^2$ respectively with the incident photo energy around $\sim$ 5 eV at the chemical potential of E$_f$ = 0 eV which increase about 38\% and  81\% respectively at a chemical potential of E$_f$ = 1.52 eV.  The systematic evolution of the bulk Fermi surface along with the high symmetry points in three dimensional Brillouin zone reveals the enhancement of bulk photovoltaics with the chemical potential in ZnGeP$_2$.  To verify our findings,  we  further explore the distribution of bulk projected bands and surface Fermi surface distribution in the energy landscape using tight binding Hamiltonian within semi infinite slab geometry.  This shows that the augmentation of bulk photovoltaics with the chemical potential is due to the surface Fermi surface states along the high symmetry $\Gamma-Z$ direction in Brillouin zone. }}Our {\bs{ thorough and detailed}} study not only provides a deeper understanding {\bs{about}} the role of Fermi surface contribution to the bulk photovoltaic responses with chemical potential,  but also suggests ZnGeP$_2$ as an ideal candidate for optoelectronics and {\bs{bulk photovoltaics}}.
\end{abstract}

\maketitle

\section{Introduction}

\par Sustainable energy demands the development of new platforms for efficient solar energy conversion.  The Shockley-Queisser limit constrains the performance of conventional solar cells based on p-n junctions,  so alternative approaches are worth exploring.  One of the most promising alternative sources of photocurrent is the bulk photovoltaic effect (BPVE) that produces photocurrents in materials lacking inversion symmetry and  unlike conventional p-n junctions,  the generated photovoltage is above the bandgap limit \cite{pal2021bulk,  PhysRevB.74.035201, PRXEnergy.2.013006,  PhysRevB.74.035201, PhysRevB.102.121111, PhysRevMaterials.8.025001, PhysRevB.74.035201,  PhysRevB.82.235204,   PhysRevB.97.241118,  PhysRevB.103.144308,  PhysRevMaterials.4.064602,  PhysRevB.104.245122,  pal2021bulk,  PhysRevLett.109.116601,  doi:10.1126/sciadv.1501524, deJuan2017}.  BPVE is a second-order nonlinear response,  which can be decomposed into a linear photogalvanic effect (LPGE) and a circular photogalvanic effect (CPGE), respectively,  based on the linear and circular polarization state of the incident light respectively \cite{PhysRevB.61.5337,  PhysRevB.94.035117,  PhysRevB.74.035201,  PhysRevB.82.235204,   PhysRevB.97.241118,  PhysRevB.103.144308,  PhysRevMaterials.4.064602,  PhysRevB.104.245122,  pal2021bulk,  PhysRevLett.109.116601,  doi:10.1126/sciadv.1501524, deJuan2017}.  The CPGE can be directly related to the Berry curvature of the Bloch bands involved in the optical transition and proportional to the topological charge of the Weyl semimetals \cite{deJuan2017}.  Microscopically, the shift current (LPGE) originates from the shift of the wave packet of Bloch electrons during interband photoexcitation \cite{doi:10.1126/sciadv.1501524}.  {\bs{Therefore,  The shift and circular photogalvanic (CPG) current originate from the interband Berry connections and Berry curvature of the Bloch bands during optical transition respectively \cite{PhysRevB.97.241118,  PhysRevLett.109.116601,  PhysRevB.94.035117}}}.

\par Nonlinear responses {\bs{open}} new opportunities for revealing the topology and band structure geometry in condensed matter systems \cite{Orenstein2021,  PhysRevB.99.155404,  PhysRevB.104.245122}.  For example,  shift and circular photocurrents,  nonlinear susceptibilities are being {\bs{ used as probes for the exploration of the}} crystallographic orientation, band structure geometry,  grain boundaries \cite{PhysRevB.61.5337, PhysRevLett.109.116601,  PhysRevLett.109.236601,  Cook2017,  doi:10.1126/sciadv.1501524,  Carvalho2020,  Yin2014},  Hall effects in both time-reversal invariant and broken systems \cite{PhysRevLett.127.206801,  Du2019, PhysRevB.100.195117,  PhysRevB.107.L081110,PhysRevB.103.144308}.  The effects of the Fermi surface and disorder {\bs{scattering}} on nonlinear optical response have started to gain attention very recently \cite{Du2019,  PhysRevLett.121.246403,  PhysRevResearch.3.L042032, Fei2018,  PhysRevB.74.155106,  PhysRevB.97.085201,  PhysRevLett.129.227401,  PhysRevLett.124.087402,  PhysRevLett.125.227401}.  Intrinsic contribution of Fermi surface,  {\bs{steaming from the photoinduced electronic transitions on the Fermi surface,}} to the BPVE has been reported in metallic systems under illumination by polarized light \cite{PhysRevResearch.3.L042032, Fei2018}.   {\bs{The intrinsic/extrinsic  contribution of Fermi surface means whether it is independent/dependent on scattering time.}} Interplay between band geometric quantities give rises to the second harmonic generation which produces resonant peaks in optical response for topological materials due to the effect of Fermi surface \cite{PhysRevB.74.155106,  PhysRevB.97.085201,  PhysRevLett.129.227401,  PhysRevLett.124.087402}.  Recent studies have shown that subbandgap photocurrents due to band broadening in semiconductors and insulators can be used to measure finite lifetimes, in contrast to clean limit {\bs{i.e., at zero temperature}} \cite{PhysRevLett.125.227401}.

\par The fundamental requirement for a material to produce a current via the BPVE is the broken inversion symmetry which allows the asymmetric photoexcitation of charge carriers induced by electron-phonon or electron-electron scattering.  BPVE has recently drawn attention in class of materials like hybrid perovskites,  topological insulators,  Weyl semimetals \cite{PhysRevLett.116.237402,  PhysRevB.95.035134,  PhysRevLett.109.116601,  PhysRevB.106.195126,  PhysRevB.97.241118}.  Chalcopyrite semicondictors have drawn attention for its potential application as a nonlinear optical material \cite{PhysRevMaterials.4.064602}.  {\bs{ Therefore,  they provide a new test bed to study BPVE as it includes both topological and non-topological materials \cite{PhysRevLett.106.016402,  PhysRevB.106.125112}. }}ZnGeP$_2$ is a one of the chalcopyrite semicondictor {\bs {derived from zinc-blende  III-V parent compound GaP where group III-Ga is replaced by group II-Zn and group IV-Ge atoms.}}  In chalcopyrite structures,  c/a ratio ({\bs{where a, c are the lattice parameters)}} is slightly lower than 2 {\bs{ compared to its ideal zinc-blende structures.  This produces a compressive uniaxial strain which affects the Brillouin zone (BZ) of  zinc-blende structures. }}The conduction-band minimum of the mother structure of GaP at X point in the BZ is folded onto the $\Gamma$ point of the ZnGeP$_2$ \cite{PhysRevB.60.8087}.  Therefore, $\Gamma$ point in the BZ has a strong impact on optical response of chalcopyrite compounds \cite{Rud1996,  petcu1996band,  mccrae1994polarized, PhysRevB.18.7099,  PhysRevB.60.8087,  PhysRevB.30.741}.  A model was proposed based on the splittings of the low-lying conduction-band states at $\Gamma$ point in order to explain the strong peak in the photoluminescence spectra \cite{Rud1996,  petcu1996band,  mccrae1994polarized}.  {\bs{Controlled  doping of Sc atoms in ZnGeP$_2$ move the chemical potential of 0.8 eV \cite{voevodin2023electrical} above Fermi level.}}

\par In this work,  we study the optical linear and non-linear response for a chalcopyrite semiconductor ZnGeP$_2$.    {\bs{We use a recently developed multi-band approach collaborated with tight binding Wannier Hamiltonian to calculate BPVE.  Using first-principles-based extensive calculations,  we report a large bulk photovoltaics like shift and CPG current conductivities and its enhancement with chemical potential due to contribution of Fermi surface states around  $\Gamma-Z$ direction in the BZ}}.  The enhancements are 38\% and  81\% respectively for shift and CPG current conductivities  {\bs{ at chemical potential of 1.52 eV}} around the peak of responses.  The distribution of Fermi surface states in the projected energy landscape is also explored using  {\bs{tight binding Wannier Hamiltonian with}} semi-infinite slab geometry to examine microscopically the contribution Fermi surface states in the BZ on the enhancement of BPVE.  The paper is organised as follows.  In Sec.  \ref{sec_II} we describe the computational details.  In Sec.  \ref{sec_III} we discuss our results on both linear optical response and bulk photovoltaic along with the electronic structure and enhancement of BPVE with chemical potential in ZnGeP$_2$.  Finally in Sec.  \ref{sec_IV},  we conclude with future perspectives.

\section{Computational details}
\label{sec_II}

\par Chalcopyrite semiconductors crystallize into the tetragonal structure with a space group I$\bar{4}$2d.  The atomic positions of ZnGeP$_2$ are Zn (0, 0, 0), Ge (0, 0, 0.5) and P (u, 0.25, 0.125),  where {\bs{u is the internal displacement parameter for the anion i.e.,  P atoms)}}.  The anion acquires an equilibrium position closer to one pair of cations as a result of dissimilar atoms as neighbors.  The structural relaxation are done in Vienna Ab initio Simulation Package (VASP) with kpoints 12$\times$12$\times$12 \cite{PhysRevB.54.11169,  PhysRevB.47.558}.  The optimized lattice constants (a and c) and the internal structural parameter (u) are a = 5.454 Å,  c = 10.707 Å and u = 0.267 Å respectively.  The density functional theoretical calculations are performed {\bs{with the optimized lattice parameters}} using the local density approximation (LDA) method as implemented within full-potential local-orbital (FPLO) code \cite{PhysRevB.59.1743}.

\par In the next step,  we use a tight-binding model in the Wannier function basis to calculate the shift and CPG current conductivities.  The tight-binding model is obtained using maximally projected Wannier functions for the Zn-3d,  4s,  Ge-3d, 4s, 4p and P-3s, 3p orbitals in the energy range of -9.0 to 5.0 eV.  BZ was sampled by a 150$\times$150$\times$150  {\bs{k-mesh grid in BZ} with satisfactory convergence {\bs{to calculate shift and CPG current conductivities. }} This Wannier model is used {\bs{further}} to calculate the spectral density for an infinite bulk system by $k_z$-integrating bulk projected band structure.  Furthermore, a semi-infinite slab is set up to calculate spectral densities of [001]-surface.  The spectral densities for both bulk and semi-infinite slab geometry are obtained using Green’s function recursion method \cite{sancho1985highly}.

\section{Results and discussion}
\label{sec_III}

\subsection{Electronic structure}

\begin{figure} [ht] 
\centering
\includegraphics[width=0.5\textwidth,angle=0]{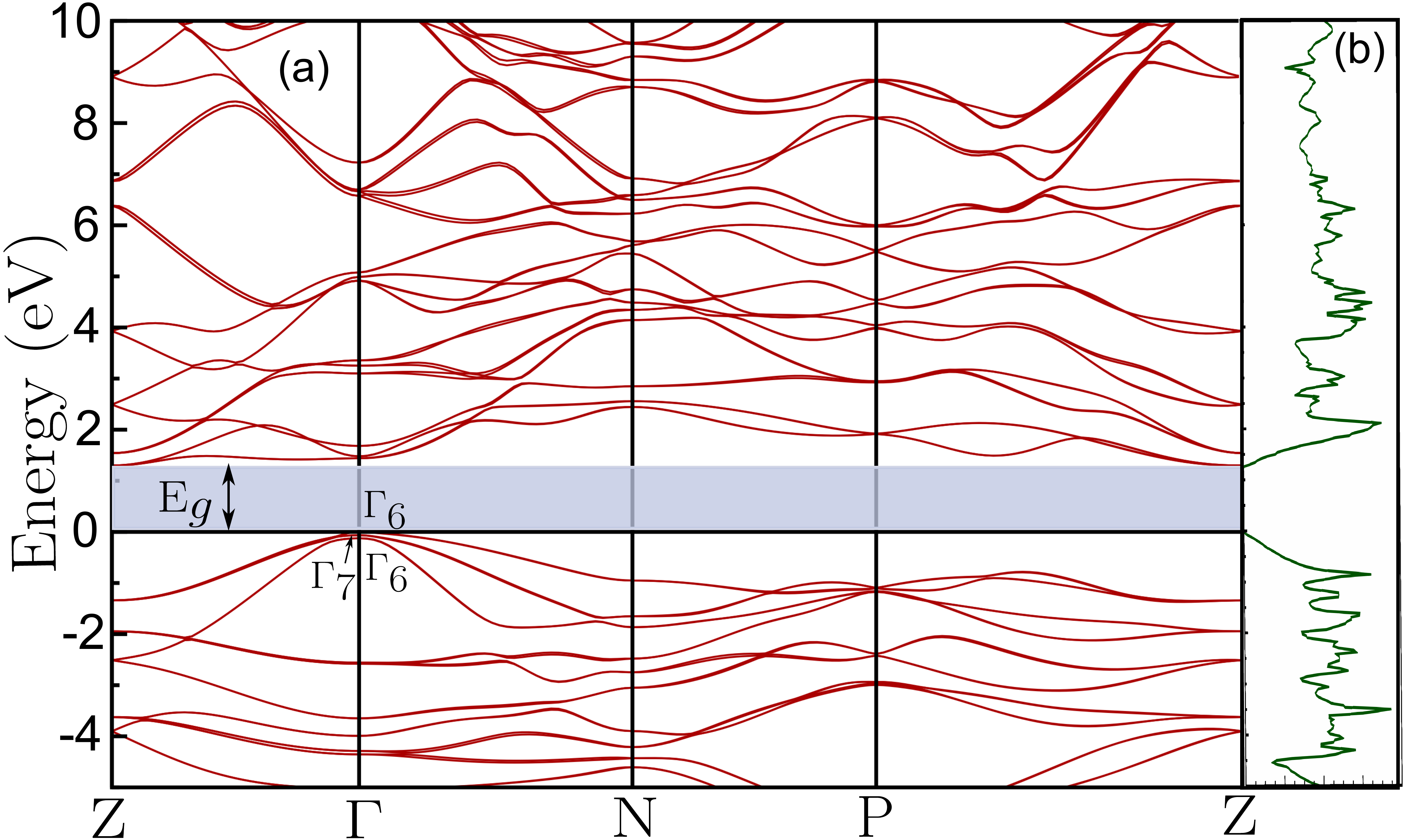} 
\caption{(a) Band structure and (b) total density of states for ZnGeP$_2$. }
\label{fig1} 
\end{figure}

\par Figure \ref{fig1} (a) represent the band structure along the high symmetry directions for ZnGeP$_2$ respectively. There is a remaining open question about the nature of its band gap whether it is direct, indirect or pseudo-direct because different studies lead to different results \cite{PhysRevB.60.8087, bendorius1972lowest, PhysRevLett.30.983,  CHIKER20043859,  zhang2015electronic}. The top of the valence band locates at $\Gamma$ point and the bottom of the conduction band minimum locates at Z point.  Therefore ZnGeP$_2$ has a indirect band gap of 1.32 eV from LSDA calculations.  In comparing the with the experimental data,  the calculated band gap is underestimated.  It is obvious that the band gap calculated by density functional theory (DFT) is smaller than the one measured experimentally \cite{shay2013ternary,  MacKinnon}.  {\bs{This error is due to the discontinuity in different exchange-correlation potentials. }}

\begin{figure*} [ht] 
\centering
\includegraphics[width=0.99\textwidth,angle=0]{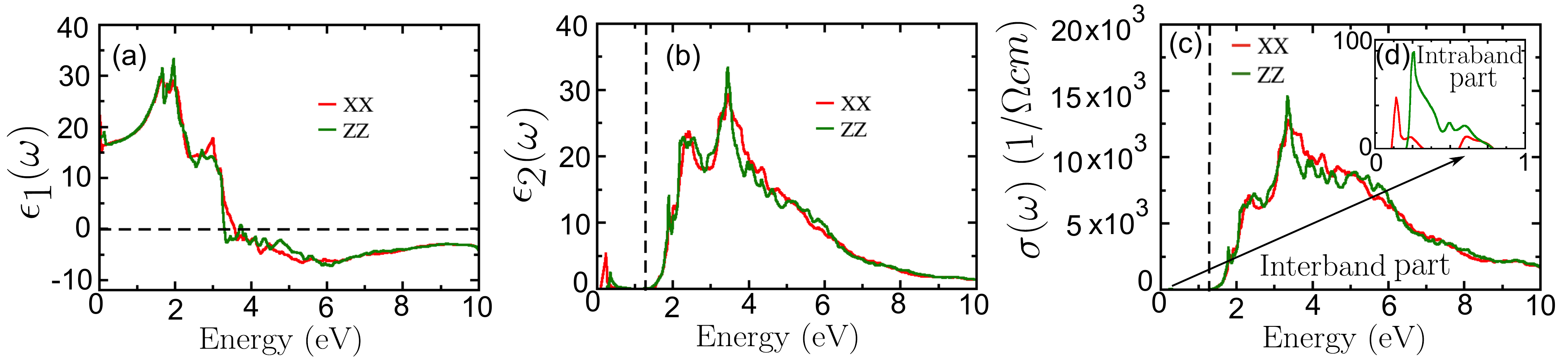} 
\caption{(a) The real and (b) imaginary parts of the dielectric constant,  and {\bs{(c)-(d) optical conductivities  as a function of the incident photon energy which show the intraband and interband parts for low and high incident photon energies respectively.}}}
\label{fig2} 
\end{figure*} 

\par Without spin-orbit coupling, triply degenerated $\Gamma_{15}$ valence band of its III–V zinc-blende compounds splits into a non-degenerate bands of $\Gamma_{4}$ and $\Gamma_{5}$ resulting a finite crystal field splitting ($\Delta_{cf}$ = E($\Gamma_{5}$)-E($\Gamma_{4}$)) in chalcopyrite structures.  In presence of spin-orbit interactions,  $\Gamma_{5}$ splits into $\Gamma_{6}$, $\Gamma_{7}$ and $\Gamma_{4} \rightarrow \Gamma_{6}$ resulting $\Delta_{cf}$ = E($\Gamma_{7}$)-E($\Gamma_{6}$) \cite{Brudny2006,  PhysRevB.60.8087}.  When the anion position parameter u of chalcopyrite crystal deviates from {\bs{its ideal value i.e.,}} 0.25,  the absolute value of crystal field splitting increases.  {\bs{This is due to the deformation of the chalcopyrite materials compared to its parent zinc-blende structures. }} Therefore $\Delta_{cf}$ is zero in GaP,   whereas it is of -102.5  meV for ZnGeP$_2$.

\par To get more insight into the the electronic structure,  we look into the total (see Fig. \ref{fig1}(b)) and partial (see Fig. \ref{sfig1} in appendix) density of states of ZnGeP$_2$. The low-energy valence bands are mainly due to P-3p state with an admixture of Ge-4p and Zn-4p states, whereas Ge-4s,  P-3d and Zn-4d states are also contributing to conduction band in addition to them.  By analyzing the partial density states,  one can clearly {\bs{observe}} the strong effects of p-d hybridization which modifies the energy gap.  The p-like states are pushed up and Zn-d like states are pushed down generating a band gap 1.32 eV for ZnGeP$_2$.

 \subsection{Linear optical response}
 
\par The {\bs{ linear optical properties}}  of ZnGeP$_2$ can be calculated from the complex dielectric function.  In the presence of an electric field,  the complex dielectric function can be divided into two parts: the intraband transition and the interband transition. In the case of metals, intraband transitions are useful, while in the case of semiconductors, interband transitions are useful.  Interband transitions {\bs{can be of two type,  direct and indirect band transitions. }} Since the indirect interband transition contributes little to dielectric function which involves electron phonon scattering, therefore it can be ignored.  By calculating the momentum matrix elements between occupied and unoccupied wave functions, the direct interband contribution can be calculated.
 
\par  The optical properties within the linear response theory are obtained from  
the imaginary part of the dielectric
function 
is given by {\bs{
\begin{align}
    \epsilon_2^{{\it{ij}}}(\omega) = {\rm Im} [\epsilon_{\it{ij}}(\omega)] 
        &=-\frac{4\pi^2 e^2} {m_0^2 \omega^2} \int d{k} \sum_{n,l} \left(f_n-f_l \right)  \nonumber \\
        & \times
            \frac{   \langle \vec{k}_n | \hat{v}_i | \vec{k}_l \rangle  \langle \vec{k}_l | \hat{v}_j | \vec{k}_n \rangle }
                {  (E_{\vec{k}_n}-E_{\vec{k}_l}-\hbar \omega -
                i\delta) }\,,
\end{align}}}
where, $i,j = (\x1,\y1,\z1)$ are the Cartesian coordinates,
$\hat{v}_i=\hat{p}_i/m_0$, $m_0$ is the free electron mass, $|{\vec k}_n
\rangle$ are the wavefunction corresponding to the band with energy $E_{{\vec k}_n}$ at
momentum ${\vec k}$ and index $n$, $f_n \equiv f(E_{\vec{k}_n})$ is the Fermi function for the
state with energy $E_{\vec{k}_n}$, and $\hbar \omega$ is the
incident photon energy. $\delta=\hbar/\tau_s$ is
the broadening parameter and depends inversely on the single particle relaxation time associated with the quantum mechanical broadening $\tau_s$.
The real part of the dielectric function can be obtained via the {\bs{Kramers-Kronig}} relation:
\begin{equation}
    \epsilon_1^{ij}(\omega) = {\rm Re} {[\epsilon_{ij}(\omega)]} = \delta_{ij}
    + \frac{1}{\pi} \mathcal{P}\int_{-\infty}^\infty \ d\omega'\
    \frac{{\rm Im} [\epsilon_{ij}(\omega')]}{\omega -\omega'} \,.
\end{equation}
The optical conductivity {{\bs{($J_i=\sigma_{ij} {\mathcal E}_j$)}} is given by : 
\begin{equation}
    \sigma_{ij}(\omega) = \frac{\omega\epsilon_2^{ij}(\omega)}{4\pi} \,.
\end{equation}
{\bs{where $J_k$ is the photocurrent generated by an electrical
field ${\mathcal E}_j$.}}

\par {\bs{Figure \ref{fig2} represents the linear optical responses in ZnGeP$_2$.  Tetragonal symmetry in ZnGeP$_2$ allows two independent linear optical components i.e one in-plane ($ij = \x1\x1 = \y1\y1$) and one out-of-plane ($ij = \z1\z1$) components.}} Figure \ref{fig2}(a)-(b) represent the real and imaginary parts of the dielectric function for ZnGeP$_2$. $\epsilon_1(\omega)$  {\bs{has}} a peak with a magnitude of 33.02 at 1.98 eV.  Then it sharply decreases between 1.98 eV and 3.36 eV and, becomes negative after that.  The minimum of $\epsilon_1(\omega)$  occurs at 6.02 eV followed by a {\bs{slow increasing trends toward zero}}.  The static dielectric constant is $\epsilon_1(0)$ = 18.01 for ZnGeP$_2$.  $\epsilon_2(\omega)$ shows that the threshold energy of the dielectric function occurs at 1.32 eV i.e above the  band gap.    The imaginary part of the dielectric constant $\epsilon_2(\omega)$ shows peaks at 2.41 and 3.44 eV respectively.  {\bs{The small contributions of  $\epsilon_2(\omega)$ below band gap originates  from the intraband transitions of optical response \cite{bhandari2020electronic,  biswas2021tunable}.  }}

\par {\bs{The calculated linear optical conductivities are presented in Fig.\ref{fig2}(c)-(d).  The total conductivity has contributions from both interband and intraband processes i.e, $\sigma(\omega) =  \sigma_{\mathrm{interband}}(\omega) + \sigma_{\mathrm{intraband}}(\omega)$.  The optical transitions from valence to conduction bands lead to interband optical conductivity above the band gap in  ZnGeP$_2$.  Therefore,  $\sigma(\omega)$ produces a strong peak at 3.34 eV followed by a weak peak at 2.32 eV due to interband optical transition as shown in Fig.\ref{fig2}(c). The intraband optical conductivity below band gap for ZnGeP$_2$ originates from Drude like conductivity due to free carriers as shown in Fig.\ref{fig2}(d). }}

 \subsection{Bulk photovoltaics and effect of Fermi surface with chemical potential}

\begin{figure*} [ht] 
\centering
\includegraphics[width=1.03\textwidth,angle=0]{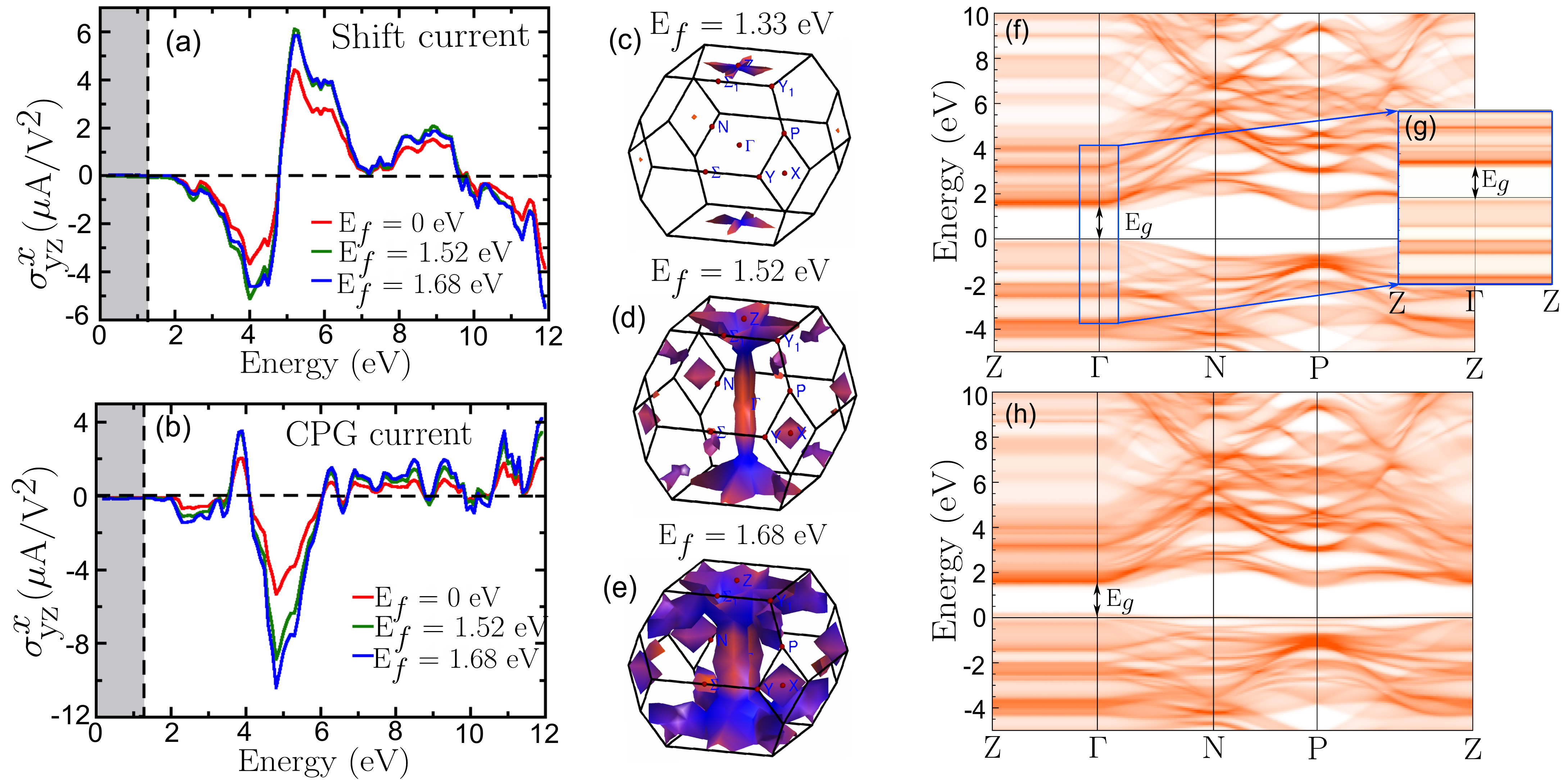} 
\caption{Fermi surface mediated enhancement of bulk photovoltaic effects for (a) shift current conductivity and (b) CPG current conductivity with different chemical potentials E$_f$.  (c)-(e) The bulk Fermi surface states in full Brillouin zone indicating high symmetry points for different {\bs{chemical potentials}}  E$_f$.  (f)-(g) $k_z$-integrated [(001)-direction] bulk projected bands. 
(h) Energy distribution curve for [001]-surface in semi infinite slab.}
\label{fig3} 
\end{figure*}

\par For the bulk photovoltaic responses,  the photoconductivity in quadratic response theory appears as
\cite{PhysRevB.19.1548,  PhysRevB.23.5590,  PhysRevB.97.241118,  PhysRevMaterials.4.064602}:
{\bs{
\begin{equation}
\begin{aligned}
     \sigma^k_{ij} &= \frac{|e|^3}{8\pi^3 \omega^2} {\rm Re} \bigg\{ \phi_{ij}
                        \sum_{\Omega=\pm \omega} \sum_{l,m,n} \int_{BZ} d{k} (f_l- f_n) \\
    & \times \frac{ \langle {\vec k}_n |  \hat{v}_i   | {\vec k}_l \rangle 
                    \langle {\vec k}_l |  \hat{v}_j   | {\vec k}_m \rangle
                    \langle {\vec k}_m |  \hat{v}_k   | {\vec k}_n \rangle}
            {(E_{\vec{k}_n}-E_{\vec{k}_m}-i\delta)(E_{\vec{k}_n}-E_{\vec{k}_l} + \hbar \Omega- i\delta)} \bigg\}.
\end{aligned}
\label{shift-eq}
\end{equation}}}
The conductivity $\sigma_{ij}^k$ ($i,j,k = \x1,\y1,\z1$) is a third rank
tensor representing the photocurrent $J_k$ generated by an electrical
field via $J_k=\sigma_{ij}^k {{\mathcal E}^*_i}{\mathcal E}_j$.  $\phi_{ij}$ is the phase difference between the
driving fields ${\mathcal E}_i$ and ${\mathcal E}_j$. The real (imaginary) part of the integral in
Eq. (\ref{shift-eq}) describes the shift (CPG) current conductivity under
linearly (circularly) polarized light respectively.

\par Now we study the non-linear photocurrent responses under linearly and circularly polarized light respectively for different chemical potentials.  As the photocurrent response arises from both real and virtual band transitions, it is generally strongly dependent on the incident photon energy.  As we consider the relaxation time approximation, therefore we used broadening parameter $\delta$ = 10 meV in our calculations.  {\bs{Chalcopyrite semiconductors belong to $\it{D_{2d}}$ (-4$\it{m}$2) point group.  Therefore,  the second order
photoconductivity ($\sigma^k_{ij}$)  tensor has the form : 
\[
  \sigma_{ij}^k=
  \left( {\begin{array}{ccccccc}
   0 & 0 & 0 & \sigma^\x1_{\y1\z1} & 0 & 0 \\
   0 & 0 & 0 & 0 & \sigma^\y1_{\x1\z1} & 0\\
   0 & 0 & 0 & 0 & 0 & \sigma^\z1_{\x1\y1}\\
  \end{array} } \right)
\]
ZnGeP$_2$ has the mirror
reflection $M_{\x1\y1}$ in the $xy$ plane and  $M_{\x1\y1}$= $M_{\y1\x1}$. It also has the $4_2$ screw rotational symmetry about the $\z1$ axis which gives  $\sigma^{\x1}_{\y1\z1}=\sigma^{\y1}_{\x1\z1}$.  Therefore,  the two independent components in  nonlinear optical photoconductivity tensor are $\sigma{^\x1_{\y1\z1}}$ and $\sigma{^\z1_{\x1\y1}}$.  For $\sigma{^\x1_{\y1\z1}}$,  the current responses are along the $x$ direction for $\y1\z1$ polarization of light. }}

\par Figure \ref{fig3}(a) presents the calculated shift current conductivity {\bs{($\sigma{^x_{yz}}$ component of second order photoconductivity tensor)}} under linear polarization of light for different {\bs{chemical potentials} E$_f$.  The shift current conductivity get a value of -3.75 $\mu$A/V$^2$ at the incident photo energy of $\sim$ 4 eV when the chemical potential is at 0 eV.   By shifting it to E$_f$ = 1.52 eV,  $\sigma_{yz}^x$ reaches to a value to -5.19 $ \mu$A/V$^2$.  Further shifting of {\bs{chemical potential} at E$_f$ = 1.68 eV,  shift current conductivity is slightly drop down to -4.59 $\mu$A/V$^2$.   The shift current conductivity gets inverted above the incident energy of 4.8 eV and reaches to a peak value of 4.46 $\mu$A/V$^2$ at E$_f$ = 0 eV {\bs{which is an order of magnitude larger than BPVE in multiferroic material BiFeO$_3$ \cite{PhysRevLett.109.236601}. }}However,  the $\sigma_{yz}^x$ enhances to 6.17 and 5.82 $\mu$A/V$^2$ for E$_f$ = 1.52 eV and 1.68 eV respectively.  The shift current conductivity has positive magnitude for the incident energy windows of (4.8-9.7) eV and gets inverted otherwise.  {\bs{The nature of the response is same for all chemical potentials but the contributions to responses can vary significantly across the BZ for a given chemical potential E$_f$ as shown in Fig. \ref{fig3}(a). }}

\par Figure \ref{fig3}(b) shows the calculated CPG current conductivity {\bs{($\sigma{^x_{yz}}$ component of second order photoconductivity tensor)}} for different {\bs{chemical potentials} E$_f$.  The CPG current conductivity at incident photo energies near the band gap is small.  It reaches a maximum value of -1.51 $\mu$A/V$^2$ at 2.32 eV within visible energy spectrum.  CPG current conductivity is negative just above band gap and reverses its direction resulting in a peak of 2.35 $\mu$A/V$^2$  at 3.8 eV  followed by a broad peak of -5.49 $\mu$A/V$^2$ at 4.8 eV for {\bs{chemical potential} E$_f$ = 0 eV.   Shifting chemical potential has a large effects also on CPG current conductivity.   However, the CPG current conductivity reaches to  -9.95 and -10.76 $\mu$A/V$^2$ at incident photon energy of 4.8 eV  for E$_f$ = 1.52 eV and 1.68 eV respectively.   {\bs{$\sigma{^z_{xy}}$ part for second order shift and CPG current conductivities are presented in appendix (see Fig.\ref{sfig2}). However,  $\sigma{^x_{yz}}$ component contributes more than $\sigma{^z_{xy}}$ component to BPVE in ZeGeP$_2$ which means BPVE with $\y1\z1$ polarization of light is higher in ZnGeP$_2$ than $\x1\y1$ polarization of light. }}

\par Fermi surface  has an intrinsic contribution to BPVE \cite{PhysRevResearch.3.L042032}.  Both the shift and CPG current conductivities enhance by shifting the {\bs{chemical potential} above band gap to (1.52 - 1.68) eV as shown in Fig.\ref{fig3}(a)-(b).  {\bs{Both the shift and CPG current conductivities get its maximum value of 6.17 $\mu$A/V$^2$ and  -10.76 $\mu$A/V$^2$ respectively for chemical potential E$_f$ = 1.52 eV.}} To investigate the connection of photovoltaic effect to Fermi surface for ZnGeP$_2$,  we study the three dimensional (3D) bulk Fermi surface at different chemical potentials in the whole BZ.  Figure \ref{sfig2}(a)-(h) (in appendix) present the systematic evolution of the bulk 3D Fermi surface along with the high symmetry points in BZ by changing the chemical potential E$_f$ from 1.33 to 2.15 eV above the band gap.

\par When the chemical potential touches at the conduction band just above the band gap (E$_f$ = 1.33 eV), the contribution of the Fermi surface appears at high symmetry $Z$ point as shown Fig.\ref{fig3}(c).  Further shifting of the chemical potential to E$_f$ = 1.52 eV,  the bulk 3D Fermi surface appears along the high symmetry line $\Gamma-Z$ directions instead of a small contribution only at $Z$ point as shown in Fig. \ref{fig3}(d).  The appearance of the bulk 3D Fermi surface broadens at the chemical potential E$_f$ = 1.68  eV (see Fig. \ref{fig3}(e)).   This enhances the both the shift and CPG current conductivities when the chemical potential are between (1.52 - 1.68) eV for ZnGeP$_2$ due to contribution of Fermi surface states along $\Gamma-Z$ direction to non-linear photocurrent.  If we further move the chemical potential above E$_f$ = 1.68 eV,  the contribution of the bulk 3D Fermi surface in the BZ decreases along $\Gamma-Z$ direction (see \ref{sfig2} (e)-(h) in appendix) which results again decrease of both shift and CPG current conductivities in ZnGeP$_2$.

\par {{\bs{The non-linear photo current conductivities (shift and CPG current) are forbidden below the band i.e 1.32 eV.  The  peak of non-linear responses appear a few eV above the band gap and outside the visible energy window.  The underestimation of band gap in semiconductors within DFT  by the local and semi-local functionals is a common issue which could be solved under the treatment of DFT with Hubbard U approximation or using different hybrid functionals.  ZnGeP$_2$ has a band gap 2.06 eV using HSE06 hybrid functional which is close to experimental value of 2.05 eV \cite{zhang2020first, shay2013ternary,  MacKinnon}.  Now the question arises on how the calculated optical responses are justified within DFT calculations.  This band gap deficiency is also addressed by shifting the conduction bands rigidly (called scissor shift) such that the electronic band gap matches with the experimental value using a ``scissors operation" on the standard DFT bands \cite{PhysRevB.37.10159,  PhysRevLett.63.1719, PhysRevMaterials.4.064602}.  Therefore under the ``scissors operation", there will be scissor shift of 0.74 eV to DFT band gap to match with experimental value for ZnGeP$_2$.  }}

\par {\bs{By implementing the procedure of ``scissors operation",   the optical response obtained from DFT is shifted by the same amount of scissor shift while maintaining the features of standard DFT \cite{PhysRevB.72.045223,  PhysRevMaterials.4.064602}.   In a very recent work,  we showed that the underlying features of band structures,  hence optical and bulk photovoltaics responses remain same from two methods, DFT + scissor shift and DFT+U,  for another material ZnSnP$_2$ in the same chalcopyrite semiconductor series  \cite{PhysRevMaterials.4.064602}.   Only there is a rigid shift of the incident photon energy in optical and bulk photovoltaics responses above which the responses start  \cite{PhysRevMaterials.4.064602}.   It implies that under ``scissors operation" the optical and bulk photovoltaics responses (DFT+scissor shift) retains the same features obtained from DFT with hybrid functional.  The qualitative features of optical responses from both DFT+scissor shift and DFT with hybrid functional remain unchanged for ZnGeP$_2$.}}

\par To investigate further,  we also study surface Fermi surface states distribution in the projected energy landscape using tight binding Wannier Hamiltonian along with the momentum space 3D BZ.  Figure \ref{fig3}(f)-(g) and (h) present the $k_z$-integrated projected band in bulk and semi-infinite slab geometry respectively.  The surface Fermi surface states available along $Z \leftarrow \Gamma \rightarrow Z$ direction at chemical potential around E$_f$ = 1.52 eV above the band gap  is due to the projection of bulk 3D Fermi surface (see Fig. \ref{fig3} (d)-(e)) which contributes in the enhancement of BPVE in ZnGeP$_2$.  However, these states decease after E$_f$ = 1.7 eV.

\section{Conclusion}
\label{sec_IV}

\par Chalcopyrite semiconductors provide a promising platform for observing bulk photovoltaics responses in addition to the linear response.  We study the Fermi surface contribution to nonlinear DC photocurrent,  namely shift and CPG current.  We find that the polarized light induces electronic transitions on the Fermi surface that contribute to the enhancement of BPVE in ZnGeP$_2$. The shift current and CPG current conductivities are 4.46 and -5.49 $\mu$A/V$^2$ with the incident photo energy at 4.8 eV for the chemical potential E$_f$ = 0 eV which produces conductivities  of 6.17 and -9.95 $\mu$A/V$^2$ respectively by shifting the chemical potentials E$_f$ to 1.52 eV.  We report the enhancement of shift and CPG current conductivities are about 38\% and  81\% due to Fermi surface states along the high symmetry $\Gamma-Z$ direction in momentum space BZ.

\par Our study is further collaborated by searching the surface Fermi surface states in energy distribution landscape which produces intrinsic contribution to BPVE.  In addition to non-linear responses, we also study the linear optical responses in ZnGeP$_2$.  Optical conductivity produces a peak of 14.83$\times10^3$ ${\Omega.cm}^{-1}$ at the incident photo energy of 3.35 eV.  Relying on these in-depth understandings of the role of Fermi surface effects and the prediction of enhancement of BPVE due to surface Fermi surface states,  ZnGeP$_2$ appears as a promising candidate for optoelectronic applications based on bulk photovoltaics.  {\bs{Our study also shed light on the root of enhanced efficiency for BPVE in noncentrosymmetric materials that could be controlled using chemical potentials. }}We believe that the results {\bs{presented}} in our work will serve as a guide for both theory and experiment in the development and optimization of the next generation bulk photovoltaics using chalcopyrite materials \cite{PhysRevMaterials.4.064602}.

\section{Acknowledgement}
BS acknowledges Department of Science and Technology,  Government of India, for financial support with reference no DST/WISE-PDF/PM-4/2023 under WISE Post-Doctoral Fellowship programme to carry out this work.  We acknowledge IFW Dresden cluster and Ulrike Nitzsche for technical support.

\bibliography{ZGP}{}

\section{Appendix}

Figure \ref{sfig1} represents the orbital projected partials density of states for different atoms for ZnGeP$_2$.  {\bs{Figure \ref{sfig2} represents $\sigma{^\z1_{\x1\y1}}$ component of shift and CPG current conductivities for different chemical potentials E$_f$.  The peak of response for the shift current and CPG current conductivities are -1.69 and -0.92 $\mu$A/V$^2$ at the chemical potential of E$_f$ = 0 which are -2.88 and -2.38 $\mu$A/V$^2$ at the chemical potential of E$_f$ = 1.52 eV. }}Figure \ref{sfig3}(a)-(h) show the calculated bulk 3D Fermi surface distribution in full Brillouin zone for different chemical potentials E$_f$ = 1.33,  1.43,  1.52,  1.68,  2.1,  2.11,  2.13, 2.15 eV respectively.

\begin{figure} [ht] 
\centering
\includegraphics[width=0.4\textwidth,angle=0]{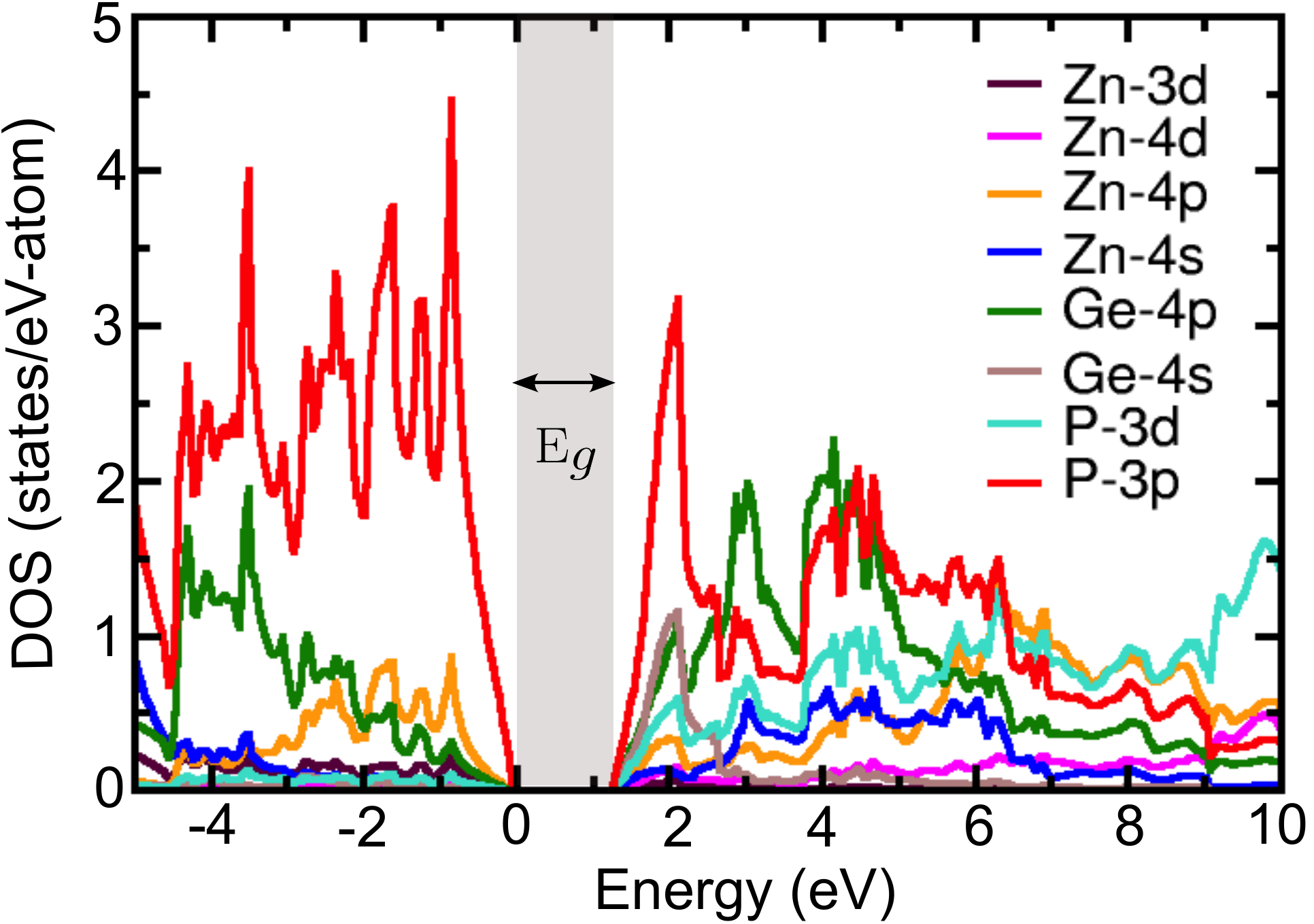} 
\caption{Orbital projected partial density of states for different atoms in ZnGeP$_2$.}
\label{sfig1} 
\end{figure}

\begin{figure*} [ht] 
\centering
\includegraphics[width=0.85\textwidth,angle=0]{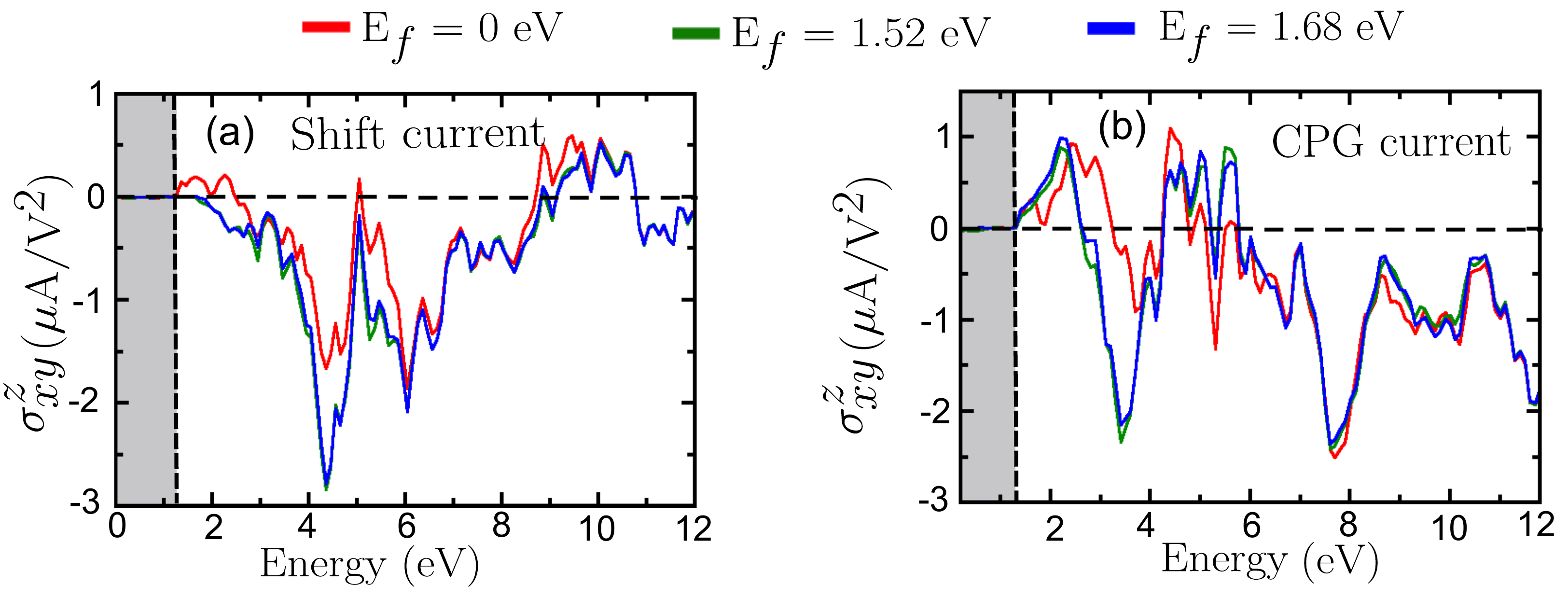} 
\caption{ {\bs{$\sigma{^\z1_{\x1\y1}}$ component of (a) shift current conductivities and (b) CPG current conductivities for different chemical potentials E$_f$.}}}
\label{sfig2} 
\end{figure*} 
  
\begin{figure*} [ht] 
\centering
\includegraphics[width=0.95\textwidth,angle=0]{SFig3.pdf} 
\caption{(a)-(h) Calculated 3D Fermi surface for ZnGeP$_2$ in full Brillouin zone with different chemical potentials E$_f$.}
\label{sfig3} 
\end{figure*}

\end{document}